\documentclass[preprint,aps,prd,showpacs,superscriptaddress,nofootinbib,tightenlines]{revtex4}

\usepackage{amsfonts}
\usepackage{multirow}
\usepackage{mathrsfs}
\usepackage{graphicx}
\usepackage{amsmath}
\usepackage{amssymb}
\usepackage{bm}
\usepackage{bbm}
\usepackage{color}
\usepackage{array}

\def\OMIT#1{}

\newcommand{\nn}{\nonumber}

\newcommand{\beq}{\begin{equation}}
\newcommand{\eeq}{\end{equation}}
\newcommand{\bqa}{\begin{eqnarray}}
\newcommand{\eqa}{\end{eqnarray}}

\begin{document}
\title{\mbox{}\\[10pt]
Next-to-leading-order QCD corrections to gluon fragmentation into
${}^1S_0^{(1,8)}$ quarkonia
}

\author{Feng Feng\footnote{F.Feng@outlook.com}}
\affiliation{China University of Mining and Technology, Beijing 100083, China\vspace{0.2cm}}
\affiliation{Institute of High Energy Physics, Chinese Academy of
Sciences, Beijing 100049, China\vspace{0.2cm}}

\author{Yu Jia\footnote{jiay@ihep.ac.cn}}
\affiliation{Institute of High Energy Physics, Chinese Academy of
Sciences, Beijing 100049, China\vspace{0.2cm}}
\affiliation{School of Physics, University of Chinese Academy of Sciences,
Beijing 100049, China\vspace{0.2cm}}

\date{\today}
\begin{abstract}
Within the NRQCD factorization framework, we compute the next-to-leading-order QCD corrections to the gluon
fragmentation into the ${}^1S_0^{(1,8)}$ Fock components of a quarkonium, at the lowest order in velocity expansion.
We follow the operator definition of the fragmentation function advanced by Collins and Soper.
The key technique underpinning our calculation is the sector decomposition method
widely used in the area of multi-loop computation. It is found that the NLO QCD corrections
have significant effects, and qualitatively modify the profiles of the corresponding
leading-order fragmentation functions.
\end{abstract}

\maketitle

Fragmentation functions (FFs) encode the essential information about the nonperturbative hadronization mechanism.
According to QCD factorization theorem~\cite{Collins:1989gx},
in a high-energy hadron collision experiment, the inclusive production rate of an identified hadron $H$ at large transverse momentum,
is dominated by the fragmentation mechanism:
\beq
d\sigma[A+B\to H(P_\perp)+X]  = \sum_i d{\hat\sigma}[A+B\to i(P_\perp/z)+X] \otimes
D_{i \to H}(z,\mu)+{\mathcal O}(1/P_\perp^2),
\label{QCD:factorization:theorem}
\eeq
where $A$, $B$ represent two colliding hadrons,
$d\hat{\sigma}$ denotes the partonic cross section,
the function $D_{i \to H}(z)$ characterizes the fragmentation probability for the parton $i$ to hadronize into a multi-hadron
state that contains the specified hadron $H$ carrying the fractional light-cone momentum $z$ with respect to the parent parton.
The sum in (\ref{QCD:factorization:theorem}) is extended over all parton specifies ($i=q,\bar{q},g$).
Similar to PDFs, FFs are also nonperturbative yet universal objects, whose scale dependence is governed by the
celebrated Dokshitzer-Gribov-Lipatov-Altarelli-Parisi (DGLAP) equation:
\beq
{d \over d \ln \mu^2} D_{g \to H}(z,\mu) =  \sum_{i} \int_z^1 {d\xi\over \xi}
P_{i g} (\xi, \alpha_s(\mu)) D_{i \to H} \left({z\over \xi},\mu\right),
\label{DGLAP:evolution}
\eeq
where we have taken the gluon fragmentation function as an explicit example, 
with $P_{ig}(\xi)$ the corresponding splitting kernel, 
and $\mu$ is usually referred to as the QCD factorization scale.
The $\mu$ dependence of the fragmentation function is such that to
compensate the $\mu$ dependence of $d\hat{\sigma}$ in (\ref{QCD:factorization:theorem}), so that the
physical production rate does not depend on this artificial scale.
Once the FF is determined at some initial scale $\mu_0$ by some means,
one can deduce its form at other scale $\mu$ by solving the evolution equation
(\ref{DGLAP:evolution}).

Unlike fragmentation into the light hadrons, the fragmentation function for a parton to a heavy quarkonium
is not necessarily a truly nonperturbative object.
Owing to the weak QCD coupling at the length scale $\sim 1/m$ ($m$ represents the heavy quark mass)
as well as the nonrelativistic nature of heavy quarkonium,
the nonrelativistic QCD (NRQCD) factorization approach~\cite{Bodwin:1994jh} may be invoked to
refactorize the quarkonium FFs as the sum of products of short-distance coefficients (SDCs)
and long-distance yet universal
NRQCD matrix elements~\cite{Braaten:1993mp,Braaten:1993rw}.
To some extent, the profiles of the quarkonia FFs are solely determined by perturbative QCD,
which renders the NRQCD approach a particularly predictive theoretical framework.
Recently, equipped with the knowledge about various fragmentation functions evaluated in this fashion,
some phenomenological predictions based on (\ref{QCD:factorization:theorem})
have been made to account for the large-$P_\perp$ $J/\psi$, $\chi_{cJ}$ and $\psi^\prime$ data samples
collected at \textsf{LHC} experiments~\cite{Bodwin:2014gia,Bodwin:2015iua}.

The original computation of FFs for quark/gluon fragmentation into the $S$-wave quarkonium
was initiated by Braaten and collaborators using NRQCD approach~\cite{Braaten:1993mp,Braaten:1993rw}.
Since then, a number of fragmentation functions for quark/gluon into various quarkonium states,
including $P$- and $D$-wave quarkonia, have been calculated in NRQCD approach
during the past two decades (for an incomplete list, see~\cite{Braaten:1993jn,Ma:1994zt,Braaten:1994kd,
Cho:1994qp,Ma:1995ci,Ma:1995vi,Braaten:1995cj,
Cheung:1995ir,Qiao:1997wb,Braaten:2000pc,Bodwin:2003wh,Sang:2009zz,Bodwin:2012xc,
Artoisenet:2014lpa,Bodwin:2014bia,Gao:2016ihc,Sepahvand:2017gup,Zhang:2017xoj,Feng:2017cjk};
for a recent compilation of various quarkonia FFs, see Ref.~\cite{Ma:2013yla}).

Most SDCs associated with quarkonium FFs were known only at LO in $\alpha_s$,
except for the simplest $g\to {}^3S_1^{(8)}$ channel~\cite{Braaten:2000pc}.
In 2014 the gluon fragmentation into the pseudoscalar (${}^1S_0^{(1)}$) quarkonium,
has been computed to next-to-leading-order (NLO) in $\alpha_s$ by Artoisenet and Braaten~\cite{Artoisenet:2014lpa}.
By that time, this is a rather challenging calculation,
where the authors have employed some complicated subtraction technique to disentangle ubiquitous IR divergences.
On the other hand, the NLO QCD correction to an analogous FF, {\it i.e.}, gluon fragmentation into the
$S$-wave spin-singlet color-octet Fock component of a charmonium, $g\to c\bar{c}({}^1S_0^{(8)}$),
remains unknown to date. The calculational challenge is expected to be comparable with \cite{Artoisenet:2014lpa}.
While the $g\to {}^1S_0^{(1)}$ fragmentation is useful for $\eta_{c,b}$ production at large $P_\perp$,
the knowledge about $g\to {}^1S_0^{(8)}$ fragmentation function
would be essential to augment our understanding about $h_{c,b}$ production
at large $P_\perp$~\cite{Feng:2017cjk}.

The goal of this work is to evaluate the NLO QCD corrections to the fragmentation functions associated with both the
gluon-to-${}^1S_0^{(1,8)}$ quarkonium, yet at lowest order in velocity expansion. We will invoke some modern techniques
widely used in the area of automated multi-loop computation, which are presumably much simpler
than that used in \cite{Artoisenet:2014lpa}.

According to the NRQCD factorization theorem~\cite{Bodwin:1994jh},
the gluon fragmentation function into charmonium $H$ can be expressed as
\beq
D_{g \to H}(z,\mu)  =   {d_1(z,\mu)\over m^3} \langle 0\vert \mathcal{O}_1^{H}(^1S_0)\vert 0 \rangle +   {d_8(z,\mu)\over m^3}
\langle 0\vert \mathcal{O}_8^{H}(^1S_0)\vert 0 \rangle+\cdots,
\label{FF:NRQCD:factorization}
\eeq
where  $d_1(z,\mu)$ and $d_8(z,\mu)$ are the desired SDCs, and the corresponding NRQCD production operators are defined by
\begin{subequations}
\label{NRQCD:production:operators}
\bqa
\mathcal{O}_1^{H}(^1S_0)  &=&
\chi^\dagger \psi  \sum_{X} |H+X\rangle  \langle H+X|
\psi^\dagger \chi,
\\
\mathcal{O}_8^{H}(^1S_0)  &=&
\chi^\dagger T^a \psi  \sum_{X} |H+X\rangle \langle H+X|
\psi^\dagger T^a \chi,
\eqa
\end{subequations}
where $T^a$ ($a=1,\ldots, N_c^2-1$) represents the generators of
$SU(N_c)$ group in fundamental representation.

The color-singlet and octet SDCs  $d_{1,8}$ can be organized in an expansion in $\alpha_s$:
\bqa
d_{1,8}(z,\mu) = d_{1,8}^{\rm LO}(z,\mu)  + \frac{\alpha_s(\mu)}{\pi} d_{1,8}^{\rm NLO}(z,\mu) + \cdots.
\eqa
These SDCs at LO in $\alpha_s$ are well-known~\cite{Braaten:1993rw,Hao:2009fa}:
\begin{subequations}
\bqa
d^{\rm LO}_1(z,\mu) &=& \frac{\alpha_s^2}{2N_c^2} \Big[ (1-z)\ln(1-z) + \frac{3}{2}z -z^2 \Big],
\\
d^{\rm LO}_8(z,\mu) &=& \frac{\alpha_s^2}{2}\frac{N_c^2-4}{N_c(N_c^2-1)} \Big[ (1-z)\ln(1-z) + \frac{3}{2}z -z^2 \Big],
\eqa
\end{subequations}
where $N_c=3$ is the number of colors.

We choose to evaluate the gluon fragmentation function in a Lorentz frame
such that the $H$ has vanishing transverse momentum.
It is customary to adopt the light-cone coordinates in calculating FF.
Any four-vector $A^\mu=(A^0,A^1,A^2,A^3)$ can be recast in the light-cone format
$A^\mu=(A^+,A^-, {\bf A}_\perp)$,
with $A^\pm \equiv {1\over \sqrt{2}}(A^0\pm A^3)$ and ${\bf A}_\perp \equiv (A^1,A^2)$.
The scalar product of two four-vectors $A$ and $B$ then becomes
$A\cdot B=A^+B^- + A^-B^+ - {\bf A}_\perp \cdot {\bf B}_\perp$.

To compute the NLO radiative correction to $d_{1,8}(z)$, let us specialize to the
gauge-invariant operator definition for the fragmentation functions as coined by Collins and Soper long ago~\cite{Collins:1981uw}.
Note that this definition was first employed by Ma to compute the quarkonium FFs in NRQCD approach~\cite{Ma:1994zt}.
For the desired $g$-to-$H$ fragmentation function, we start from the operator definition~\cite{Collins:1981uw}
(also see \cite{Bodwin:2003wh,Bodwin:2012xc}):
\bqa
\label{CS:def:Fragmentation:Function}
& & D_{g \to H}(z,\mu) =
\frac{-g_{\mu \nu}z^{D-3} }{ 2\pi k^+ (N_c^2-1)(D-2) }
\int_{-\infty}^{+\infty} \!dx^- \, e^{-i k^+ x^-}
\\
&& \times
\langle 0 | G^{+\mu}_c(0)
\Phi^\dagger(0,0,{\bf 0}_\perp)_{cb}  \sum_{X} |H(P)+X\rangle \langle H(P)+X|
\Phi(0,x^-,{\bf 0}_\perp)_{ba} G^{+\nu}_a(0,x^-,{\bf 0}_\perp) \vert 0 \rangle,
\nn
\eqa
where $z$ denotes the fraction of the $+$-momentum carried by $H$ with respect to the gluon,
$D=4-2\epsilon$ signifies the space-time dimensions,
$G_{\mu\nu}$ is the matrix-valued gluon field-strength tensor in the adjoint representation of $SU(N_c)$,
$k^+ = P^+/z$ is the $+$-component momentum of injected by the gluon field strength operator.
$\mu$ is the renormalization scale for this composite nonlocal operator.
The insertion of the intermediate states implies
that in the asymptotic future, one only needs project out those out states
that contain a charmonium $H$ carrying the definite momentum $P^\mu$, with additional
unobserved hadrons labelled by the symbol $X$.

The gauge link (eikonal factor) $\Phi(0,x^-,{\bf 0}_\perp)$ in (\ref{CS:def:Fragmentation:Function}) is a
path-ordered exponential of the gluon field,
whose role is to ensure the gauge invariance of the FF:
\beq
\Phi(0,x^-,{\bf 0}_\perp)_{ba} = \texttt{P} \exp
\left[ i g_s \int_{x^-}^\infty d y^- n\cdot A(0^+,y^-,{\bf 0}_\perp) \right]_{ba},
\label{Gauge:Link:Definition}
\eeq
where $\texttt{P}$ implies the path-ordering, $g_s$ is the QCD coupling constant, and
$A^\mu$ designates the matrix-valued gluon field in the adjoint representation.
$n^\mu =(0,1,{\bf 0}_\perp)$ is a reference null 4-vector.

We proceed to calculate the SDCs $d_{1,8}(z)$ by the standard perturbative matching strategy,
{\it i.e.}, by replacing the physical charmonium $H$ in (\ref{FF:NRQCD:factorization})
with the free $c\bar{c}$ pair of quantum number $^1S_0^{(1,8)}$.
Computing the QCD side from
\eqref{CS:def:Fragmentation:Function} in perturbation theory,
and using the following NRQCD matrix elements,
\beq
\langle 0\vert {\cal O}^{c\bar{c}}_1(^1S_0) \vert 0 \rangle = 2N_c,
\qquad \langle 0\vert {\cal O}^{c\bar{c}}_8(^1S_0) \vert 0 \rangle = N_c^2-1,
\eeq
one can readily solve for $d_{1,8}(z)$ order by order in $\alpha_s$.

\begin{figure}[tb]
\centering
\includegraphics[width=0.8\textwidth]{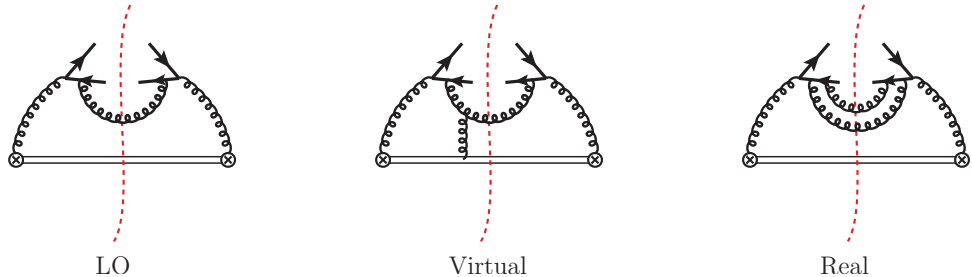}
\caption{Representative cut diagrams for the gluon fragmentation function $d_{g\to c\bar{c}(^1S_0^{(1,8)})}(z)$.
The cap represents the gluonic field strength operator $G_a^{+\nu}$, and double line signifies the eikonal line.
\label{Feynman:diagram}}
\end{figure}

Since Eq.~(\ref{CS:def:Fragmentation:Function}) is manifestly gauge-invariant, for simplicity
we specialize to the Feynman gauge.
Dimensional regularization, with spacetime dimensions $d=4-2\epsilon$,
is used throughout to regularize both UV and IR divergences.
We wrote a private \textsf{Mathematica} package to automatically generate the cut Feynman diagrams and the affiliated
amplitudes that correspond to the perturbative fragmentation function defined in \eqref{CS:def:Fragmentation:Function}.
Basing on the package \textsf{Qgraf}~\cite{Nogueira:1991ex}, we have implemented the Feynman rules for the
eikonal propagator and vertex~\cite{Collins:1981uw} as well as those for conventional QCD propagators and vertices.
Some typical Feynman diagrams for $d_{1,8}$ through NLO in $\alpha_s$ are shown in Fig.~\ref{Feynman:diagram}.

To project the $c\bar{c}$ pair onto the intended spin/orbital/color states, it is convenient to employ the
familiar covariant projector technique to expedite the calculation~\cite{Petrelli:1997ge}:
\begin{subequations}
\bqa
\Pi_1 &=& \frac{1}{\sqrt{8m^3}}\left({P\!\!\!\slash\over 2}- m\right)\gamma_5 \left({P\!\!\!\slash\over 2}+m\right) \otimes { \texttt{1}_c  \over \sqrt{N_c}},
\\
\Pi_8^a &=& \frac{1}{\sqrt{8m^3}}\left({P\!\!\!\slash\over 2}- m\right)\gamma_5 \left({P\!\!\!\slash\over 2}+m\right)  \otimes \sqrt{2} T^a,
\eqa
\label{spin:color:projectors}
\end{subequations}
where $P^\mu$ designates the total momentum of the $c\bar{c}$ pair, $\texttt{1}_c$ is the $N_c\times N_c$-dimensional
unit matrix. Since we are only interested in the LO contribution in velocity expansion,
we have neglected in \eqref{spin:color:projectors} the relative momentum between $c$ and $\bar{c}$ on both sides of the cut,
consequently $P^2=4m^2$.

With the aid of the covariant projector \eqref{spin:color:projectors}, we utilize the packages \textsf{FeynCalc/FormLink}~\cite{Mertig:1990an,Feng:2012tk}
to conduct the Dirac/color trace operation. We also use the package \textsf{Apart}~\cite{Feng:2012iq} to simplify the
amplitude by the method of partial fraction, to make the loop integration in next step easier.

A specific trait of the fragmentation function is its cut diagram structure,
which is resulting from the insertion of the asymptotic out states in (\ref{CS:def:Fragmentation:Function}).
As a consequence, the corresponding cut-line phase space integration measure reads~\cite{Bodwin:2003wh,Bodwin:2012xc}
\bqa
d\Phi_n &=& {8\pi m \over S_n} \delta(k^+-P^+-\sum_{i=1}^n k_i^+) \prod_{i=1}^n \frac{dk^+_i}{2k_i^+}\frac{d^{D-2}k_{i\perp}}{(2\pi)^{D-1}} \theta(k^+_i),\label{phase:space}
\eqa
where $k_i$ ($i=1,2$) stands for the momentum of the $i$-th on-shell gluon that pass through the cut,
and $S_n$ is the statistical factor for $n$ identical gluons~\footnote{In real correction diagrams, it is also
possible that a pair of massless quark and antiquark passes through the cut in addition to the $c\bar{c}$ pair.}.
For our purpose, suffices it to know $S_1=1$ and $S_n=2$. It is important to note that integration over $k_i^+$
can be transformed into a parametric integration in a finite interval, but the integration over the transverse momentum
$k_{i,\perp}$ are completely unbounded, {\it i.e.}, from $-\infty$ to $+\infty$.
This feature may persuade us that integration over $k_{i,\perp}$ could be regarded as
loop integration in $D-2$-dimensional spacetime.

The real correction diagrams are featured by those in Fig.~\ref{Feynman:diagram} with two gluons passing through the cut
besides the $c\bar{c}({}^1S_0^{(1,8)})$,
while the virtual correction diagrams are defined those with only one additional gluon passing through the cut.
For the former type of contribution, it has been recently shown~\cite{Zhang:2017xoj}
that, the integration-by-part (IBP) technique developed in the area of multi-loop calculation
can be effectively invoked to reduce the integrand into the linear combination of a set of
simpler master integrals, which can then be analytically ascertained.
The gluon fragmentation into $c\bar{c}({}^3S_1^{(1)})$ and $c\bar{c}({}^1P_1^{(1)})$ have been analytically
evaluated in this manner~\cite{Zhang:2017xoj,Feng:2017cjk}.

In this work, rather than utilize the IBP technique,
we apply the influential {\it sector decomposition} method~\cite{Binoth:2000ps,Binoth:2003ak}
to evaluate both real and virtual correction diagrams.
Consequently, we will present our final results in an entirely numerical fashion.
In our opinion, the approach used in this work
appears to be more amenable to automated calculation, and yield more accurate numerical predictions
than the subtraction approach adopted in \cite{Artoisenet:2014lpa}.

The first step is to combine all the propagators in a cut amplitude using Feynman parametrization.
For real correction contribution, it is straightforward to accomplish two-loop integration over
$k_{1,2\,\perp}$ in $D-2$-dimensional spacetime.
We are then left with multi-fold integrals over Feynman parameters,
which is ready and suitable for conducting sector decomposition
with the help of the package \textsf{FIESTA}~\cite{Smirnov:2013eza}.
For virtual correction diagram, the situation is somewhat more subtle.
One cannot carry out the integration over loop momentum $l$
in $D$-dimensional spacetime and the transverse momentum $k_{1,\perp}$ in
$D-2$-dimensional spacetime simultaneously.
The key is to first integrate over $l$ by the standard Gaussian method,
and the resulting expression is still of quadratic form with respect to $k_{1,\perp}$,
so we can continue to integrate over $k_{1,\perp}$ using Gaussian method,
and end up with multi-fold integrals over Feynman parameters.
This form is again suitable for conducting sector decomposition with the aid of
\textsf{FIESTA}~\cite{Smirnov:2013eza}.
The role of sector decomposition method~\cite{Binoth:2000ps,Binoth:2003ak} is to
disentangle various poles, typically with many finite multi-variable parametric integrals
as output for the corresponding coefficients.
We finally adopt the powerful integrators {\tt CubPack}~\cite{CubPack} and {\tt ParInt}~\cite{parint} to
carry out those numerical integration to high precision.

Adding both real and virtual correction pieces, and implementing the contribution from  counterterm QCD lagrangian
(we renormalize the QCD coupling constant according to the $\overline{\rm MS}$ scheme),
we find that the NLO SDCs in both
color-singlet and octet channels are absent of IR pole,
but still contain an extra single UV pole, whose coefficients are dependent
on the momentum fraction $z$.
This indicates that the fragmentation function still requires an additional
operator renormalization~\cite{Collins:1981uw,Bodwin:2014bia}:
\beq
D^{\overline{\rm MS}}_{g\to H}(z,\mu) = D_{g\to H}(z,\mu) -
{1\over \epsilon}{\alpha_s\over 2\pi} \int_z^1 \!\!{dy\over y}\, P_{gg}(y) D_{g\to H}(z/y,\mu),
\label{DGLAP:renormalization}
\eeq
where $P_{gg}(y)$ represents the Altarelli-Parisi splitting kernel for $g\to g$:
\beq
P_{gg}(y) = 2N_c \left[ \frac{y}{(1-y)_+} + \frac{1-y}{y} + y(1-y) \right] + \beta_0 \delta(1-y),
\eeq
with $\beta_0=(11 N_c-2 n_f)/6$ the one-loop QCD $\beta$ function, and $n_f=n_L+n_H$ signifies the number of active flavors.
Here $n_L$ denotes the number of light quarks, and $n_H$ denotes the number of heavy quarks comprising the quarkonium.
Note in \eqref{DGLAP:renormalization} the UV pole is subtracted in accordance with the $\overline{\rm MS}$ procedure.

Following the DGLAP renormalization procedure specified in
\eqref{DGLAP:renormalization}, we then extract the intended finite
SDCs $d^{\rm NLO}_{1,8}(z,\mu)$ through NLO in $\alpha_s$.
It is convenient to divide them into several parts:
\beq
d^{\rm NLO}_{1,8}(z,\mu) = c^{(1,8)}_0(z) \ln\frac{\mu}{m}  + \alpha_s^2 \left[ c^{(1,8)}_1(z) + n_L c^{(1,8)}_2(z)  + n_H c^{(1,8)}_3(z) \right].
\label{Dividing:dNLO:several:parts}
\eeq
For clarity, we have separated the light-quark contributions from heavy quark.

The coefficient function of $\ln \mu$ can be analytically deduced:
\bqa
\label{c0:analytical:expression}
c^{(1,8)}_0(z) & \equiv & \int_{z}^1 \frac{dy}{y} \Big[ P_{gg}(y) + 2\beta_0 \delta(1-y) \Big] d_{1,8}^{\rm LO}(z/y)
\nn \\
&=& 3\beta_0 d_{1,8}^{\rm LO}(z) + \frac{\alpha_s^2}{6z} {\mathcal F}^{(1,8)} \bigg\{
\left(-6 z^2-12 z\right) \text{Li}_2(z)+\left(6 z-6 z^2\right) \ln^2(1-z)
\nn\\
&+& \left(6 z^2-6 z\right) \ln(1-z) \ln z+\left(-9 z^3+6 z^2+3 z+3\right) \ln(1-z)
\\
&+& \left(3 z^3-12 z^2\right) \ln z +\pi ^2 \left(2 z^2+z\right) +1 -9 z +17 z^3-9 z^2
\bigg\},
\nn
\eqa
with the color factors
\beq
{\mathcal F}^{(1)} =\frac{1}{N_c},\qquad {\mathcal F}^{(8)} = \frac{N_c^2-4}{N_c^2-1}.
\eeq
We notice that \eqref{c0:analytical:expression} diverges as $1/z$ in the $z\to 0$ limit.

It is impossible for our approach to deduce the analytical expressions for those non-logarithmic coefficient functions 
$c^{(1,8)}_{i}(z)$ ($i=1,2,3$).
Nevertheless, for a given $z$, we can compute their numerical values to very high precision, within relatively short time.
For reader's convenience, we have tabulated in Table~{\ref{c1-table}} and Table~{\ref{c8-table}} 
the values of $c^{(1,8)}_{1,2,3}(z)$ for a number of representative values of $z$.

\begin{table}[t]
\caption{\label{c1-table}
Numerical values of non-logarithmic color-singlet coefficient functions $c_{1,2,3}^{(1)}(z)$
as introduced in \eqref{Dividing:dNLO:several:parts}.
We caution that the actual values of $c^{(1)}_2(z)$ and $c^{(1)}_3(z)$ should be multiplied by a factor $10^{-2}$.
}
\newcolumntype{L}{>{$}l<{$}}
\newcolumntype{R}{>{$}r<{$}}
\newcolumntype{C}{>{$}c<{$}}
\begin{ruledtabular}
\begin{tabular}{CRRRCRRR}
 z & c^{(1)}_1(z) & c^{(1)}_2(z) & c^{(1)}_3(z) & z & c^{(1)}_1(z) & c^{(1)}_2(z) & c^{(1)}_3(z)  \\
\hline
 0.05 & -1.2635(2) & 0.1932(2) & 0.16065(8) & 0.55 & 0.17598(9) & 0.0825(2) & 0.18142(8) \\
 0.10 & -0.46478(3) & 0.2744(2) & 0.24159(7) & 0.60 & 0.1771(1) & 0.0454(2) & 0.15398(8) \\
 0.15 & -0.19605(4) & 0.3106(2) & 0.28783(7) & 0.65 & 0.17507(8) & 0.0119(2) & 0.12961(8) \\
 0.20 & -0.05978(6) & 0.3180(2) & 0.31024(7) & 0.70 & 0.17039(7) & -0.0187(2) & 0.10914(8) \\
 0.25 & 0.02256(5) & 0.3059(2) & 0.31542(7) & 0.75 & 0.1624(1) & -0.0487(1) & 0.09259(8) \\
 0.30 & 0.07698(6) & 0.2805(2) & 0.30781(7) & 0.80 & 0.1493(1) & -0.0845(1) & 0.07856(8) \\
 0.35 & 0.11772(9) & 0.2462(2) & 0.29084(7) & 0.85 & 0.12909(9) & -0.1399(1) & 0.06286(8) \\
 0.40 & 0.14046(5) & 0.2067(2) & 0.26738(7) & 0.90 & 0.09093(9) & -0.2499(1) & 0.03489(8) \\
 0.45 & 0.15814(5) & 0.1647(2) & 0.23991(8) & 0.95 & -0.0102(1) & -0.5280(1) & -0.03512(8) \\
 0.50 & 0.1694(1) & 0.1227(2) & 0.21061(8) & 0.99 & -0.4349(7) & -1.3903(1) & -0.20468(8) \\
\end{tabular}
\end{ruledtabular}
\end{table}

\begin{table}[t]
\caption{\label{c8-table}
Numerical values of non-logarithmic color-octet coefficient functions $c_{1,2,3}^{(8)}(z)$
defined in \eqref{Dividing:dNLO:several:parts}.
We caution that the actual values of $c^{(8)}_2(z)$ and $c^{(8)}_3(z)$ should be multiplied by a factor $10^{-2}$.
}
\newcolumntype{L}{>{$}l<{$}}
\newcolumntype{R}{>{$}r<{$}}
\newcolumntype{C}{>{$}c<{$}}
\begin{ruledtabular}
\begin{tabular}{CRRRCRRR}
 z & c^{(8)}_1(z) & c^{(8)}_2(z) & c^{(8)}_3(z) & z & c^{(8)}_1(z) & c^{(8)}_2(z) & c^{(8)}_3(z)  \\
\hline
0.02 & -6.928(3) & 0.1976(4) & 0.1568(5) & 0.55 & 0.3105(3) & 0.155(3) & 0.34016(6) \\
 0.05 & -2.4747(3) & 0.362(4) & 0.30134(6) & 0.60 & 0.3486(4) & 0.085(3) & 0.28871(6) \\
 0.10 & -1.0082(3) & 0.515(4) & 0.45336(6) & 0.65 & 0.3845(3) & 0.022(3) & 0.24302(6) \\
 0.15 & -0.5161(3) & 0.582(4) & 0.53967(6) & 0.70 & 0.4189(4) & -0.035(3) & 0.20464(7) \\
 0.20 & -0.2620(3) & 0.596(4) & 0.58170(6) & 0.75 & 0.4522(3) & -0.091(3) & 0.17361(7) \\
 0.25 & -0.1020(3) & 0.574(4) & 0.59142(6) & 0.80 & 0.4818(3) & -0.158(2) & 0.14731(7) \\
 0.30 & 0.0109(3) & 0.526(4) & 0.57714(6) & 0.85 & 0.5001(3) & -0.262(2) & 0.11787(7) \\
 0.35 & 0.0964(3) & 0.462(4) & 0.54532(6) & 0.90 & 0.4785(3) & -0.469(2) & 0.06541(7) \\
 0.40 & 0.1644(3) & 0.388(4) & 0.50134(6) & 0.95 & 0.2864(4) & -0.990(2) & -0.06585(7) \\
 0.45 & 0.2200(5) & 0.309(3) & 0.44983(6) & 0.98 & -0.2839(3) & -1.8829(1) & -0.26181(6) \\
 0.50 & 0.2681(5) & 0.230(3) & 0.3949(5) & 0.99 & -0.9373(9) & -2.6068(2) & -0.38376(6) \\
\end{tabular}
\end{ruledtabular}
\end{table}

For numerical investigation, we take $\mu$ as twice heavy quark mass,
and adopt the following input parameters~\cite{Feng:2015uha,Feng:2017hlu}:
\beq
m_c = 1.68\,\text{GeV},\; m_b=4.78\,\text{GeV},\; \alpha_s(2m_c)=0.242,\; \alpha_s(2m_b)=0.180.
\eeq
We have taken $n_L=3,4$ for charmonium and bottomonium, respectively, and sent $n_H = 0$ so $n_f=n_L$.

\begin{figure}[tb]
\centering
\includegraphics[width=0.45\textwidth]{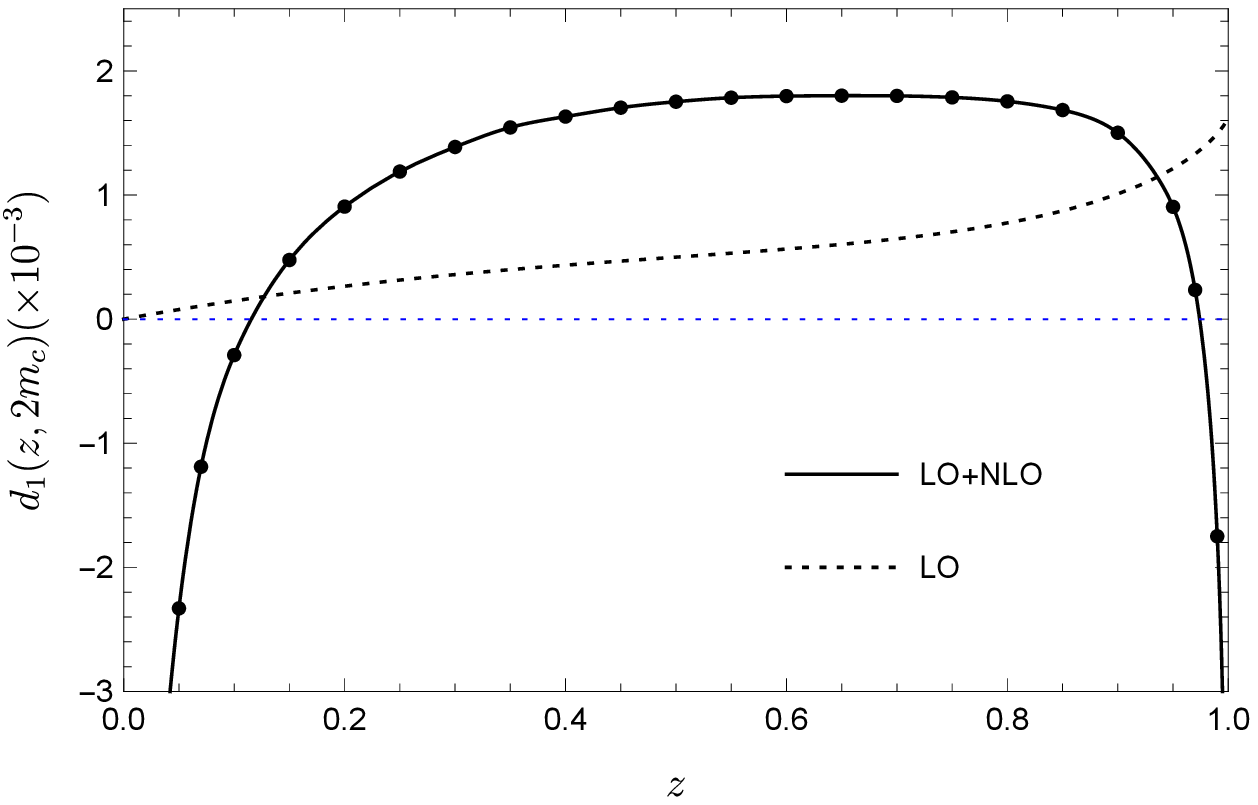} \;
\includegraphics[width=0.45\textwidth]{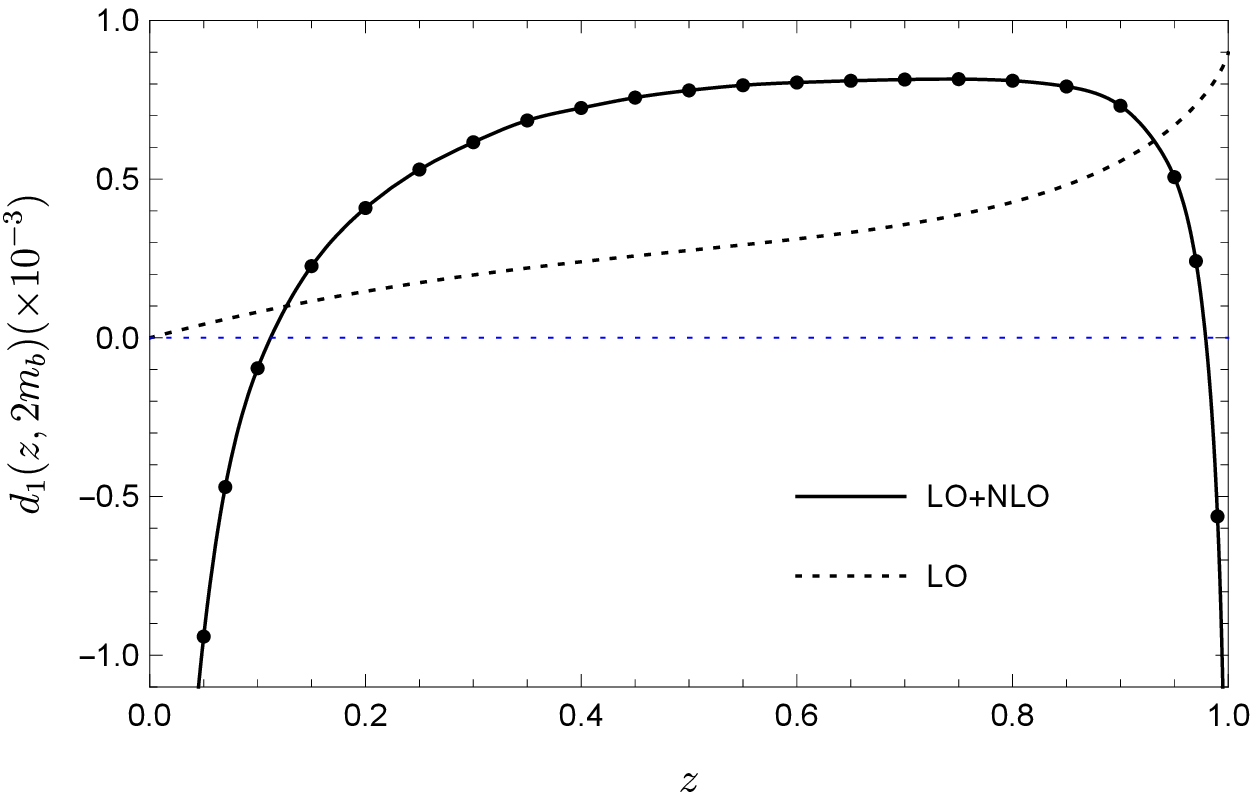}
\includegraphics[width=0.45\textwidth]{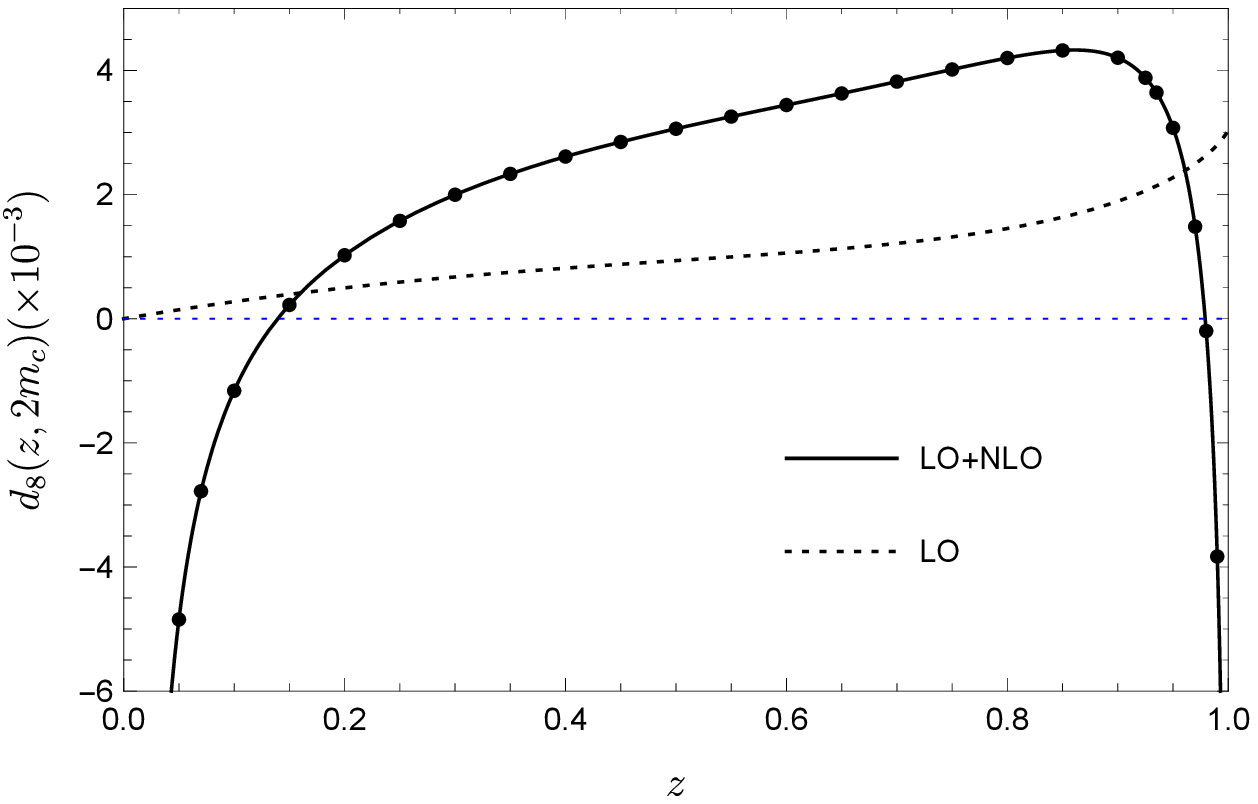} \;
\includegraphics[width=0.45\textwidth]{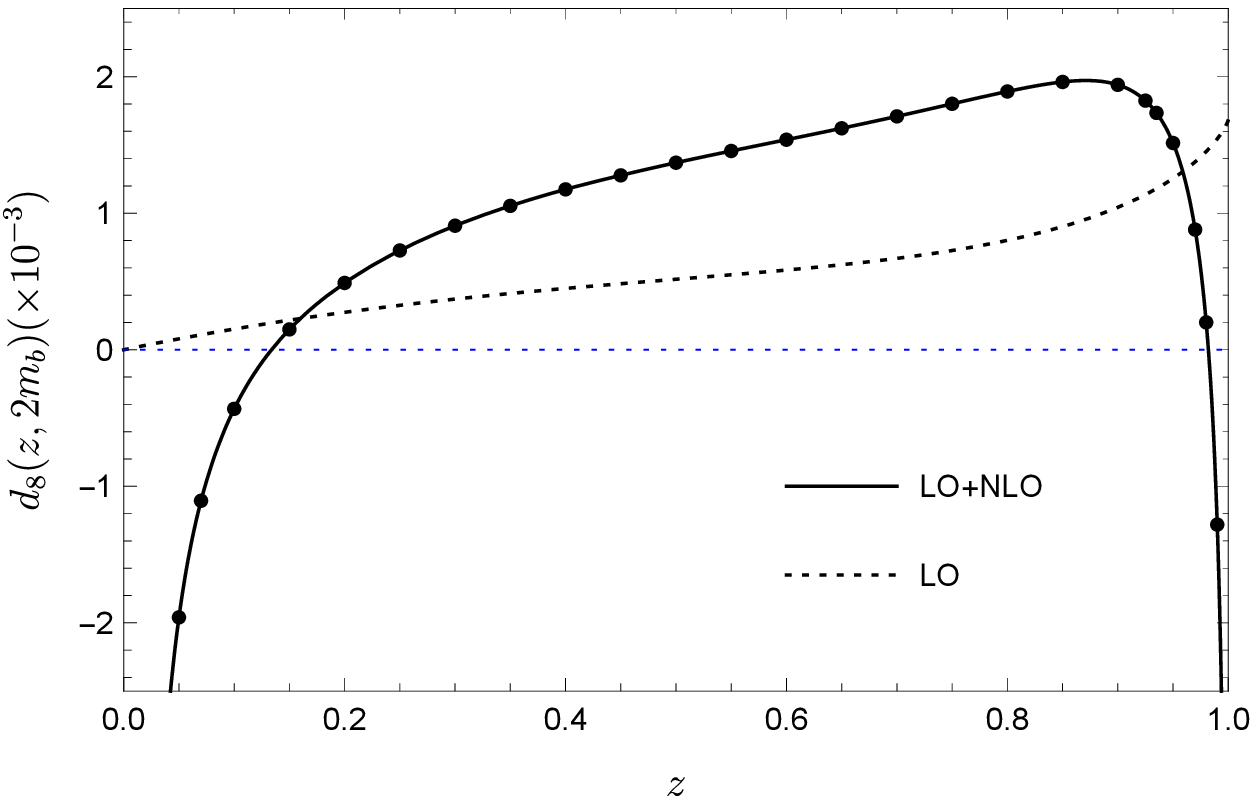}
\caption{The SDCs $d_{1,8}(z)$ associated with gluon fragmentation into quarkonium, including the NLO QCD corrections.
The two figures in the top row correspond to $d_1(z)$ for gluon fragmentation into the color-singlet charmonium and bottomonium,
while the two figures in the lower panel correspond to $d_8(z)$ for gluon into the color-octet charmonium and bottomonium.
\label{figDg}}
\end{figure}

The profiles of SDC $d_{1,8}(z)$ through the NLO in $\alpha_s$  are displayed in Fig.~\ref{figDg},
for gluon fragmentation into both charmonium and bottomonium.
Apparently, the NLO QCD corrections have a significant impact on both channels,
qualitatively changing the shape of LO fragmentation functions.
It should be mentioned that our result for $d_{1}^{\rm NLO}(z)$ disagrees with that in \cite{Artoisenet:2014lpa}, especially at low $z$ region.
Since we already knew from \eqref{c0:analytical:expression} that our $d_{1,8}^{\rm NLO}(z,\mu)$ divergences $\propto 1/z$ in
$z\to 0$ limit, we are thereby unable to present a {\it finite} prediction to the total 
fragmentation probability at NLO in $\alpha_s$.

In summary, in this work we have computed the NLO QCD corrections to the gluon
fragmentation into both ${}^1S_0^{(1,8)}$ Fock components of quarkonium,
at the LO in velocity expansion in NRQCD factorization.
It is most transparent to start from Collins and Soper's operator definition of the fragmentation function
when investigating the higher order radiative corrections.
To facilitate the numerical evaluation of virtual and real correction contributions,
we have employed an automated approach that is based crucially upon
the sector decomposition technique,
It turns out that this method is quite efficient and systematic,
and a good numerical accuracy can be achieved with modest calculational expense.
It is found that the NLO QCD corrections in both color-singlet and octet channels
have important impact, and qualitatively modify the profiles of the corresponding LO fragmentation functions.
Our results might be useful to strengthen our understanding about large-$P_\perp$ production of $\eta_{c,b}$ and $h_{c,b}$
at LHC experiment.

\vspace{0.3 cm}

\noindent{\it Note added.} While we were finalizing the manuscript,
a preprint has recently appeared, which also computes the NLO QCD corrections to the fragmentation function for
gluon-to-${}^1S_0^{(8)}$ quarkonium, yet using the FKS subtraction scheme~\cite{Artoisenet:2018dbs}.
Their numerical results appear to be compatible with ours.
We also compare our NLO radiative corrections for both $g\to {}^1S_0^{(1,8)}$ fragmentation functions
at some typical values of $z$ with a forthcoming paper~\cite{Zhang:2018},
and find perfect agreement.

\begin{acknowledgments}
The work of F.~F. is supported by the National Natural Science Foundation of China under Grant No.~11505285, No.~11875318, and by the Yue Qi Young Scholar Project in CUMTB.
The work of  Y.~J. is supported in part by the National Natural Science Foundation of China under Grants No.~11475188,
No.~11621131001 (CRC110 by DGF and NSFC).
\end{acknowledgments}


\begin{thebibliography}{99}

\bibitem{Collins:1989gx}
  J.~C.~Collins, D.~E.~Soper and G.~F.~Sterman,
  Adv.\ Ser.\ Direct.\ High Energy Phys.\  {\bf 5}, 1 (1989)
  [hep-ph/0409313].

\bibitem{Bodwin:1994jh}
  G.~T.~Bodwin, E.~Braaten and G.~P.~Lepage,
  Phys.\ Rev.\ D {\bf 51}, 1125 (1995)
  Erratum: [Phys.\ Rev.\ D {\bf 55}, 5853 (1997)]
  [hep-ph/9407339].

\bibitem{Braaten:1993mp}
  E.~Braaten, K.~m.~Cheung and T.~C.~Yuan,
  Phys.\ Rev.\ D {\bf 48}, 4230 (1993)
  [hep-ph/9302307].

\bibitem{Braaten:1993rw}
  E.~Braaten and T.~C.~Yuan,
  Phys.\ Rev.\ Lett.\  {\bf 71}, 1673 (1993)
  [hep-ph/9303205].

\bibitem{Bodwin:2014gia}
  G.~T.~Bodwin, H.~S.~Chung, U.~R.~Kim and J.~Lee,
  Phys.\ Rev.\ Lett.\  {\bf 113}, no. 2, 022001 (2014)
  [arXiv:1403.3612 [hep-ph]].

\bibitem{Bodwin:2015iua}
  G.~T.~Bodwin, K.~T.~Chao, H.~S.~Chung, U.~R.~Kim, J.~Lee and Y.~Q.~Ma,
  Phys.\ Rev.\ D {\bf 93}, no. 3, 034041 (2016)
  [arXiv:1509.07904 [hep-ph]].

\bibitem{Braaten:1993jn}
  E.~Braaten, K.~m.~Cheung and T.~C.~Yuan,
  Phys.\ Rev.\ D {\bf 48}, no. 11, R5049 (1993)
  [hep-ph/9305206].

\bibitem{Ma:1994zt}
  J.~P.~Ma,
  Phys.\ Lett.\ B {\bf 332}, 398 (1994)
  [hep-ph/9401249].

\bibitem{Braaten:1994kd}
  E.~Braaten and T.~C.~Yuan,
  Phys.\ Rev.\ D {\bf 50}, 3176 (1994)
  [hep-ph/9403401].

\bibitem{Cho:1994qp}
  P.~L.~Cho and M.~B.~Wise,
  Phys.\ Rev.\ D {\bf 51}, 3352 (1995)
  [hep-ph/9410214].

\bibitem{Ma:1995ci}
  J.~P.~Ma,
  Nucl.\ Phys.\ B {\bf 447}, 405 (1995)
  [hep-ph/9503346].

\bibitem{Ma:1995vi}
  J.~P.~Ma,
  Phys.\ Rev.\ D {\bf 53}, 1185 (1996)
  [hep-ph/9504263].

\bibitem{Braaten:1995cj}
  E.~Braaten and T.~C.~Yuan,
  Phys.\ Rev.\ D {\bf 52}, 6627 (1995)
  [hep-ph/9507398].

\bibitem{Cheung:1995ir}
  K.~m.~Cheung and T.~C.~Yuan,
  Phys.\ Rev.\ D {\bf 53}, 3591 (1996)
  [hep-ph/9510208].

\bibitem{Qiao:1997wb}
  C.~f.~Qiao, F.~Yuan and K.~T.~Chao,
  Phys.\ Rev.\ D {\bf 55}, 5437 (1997)
  [hep-ph/9701249].

\bibitem{Braaten:2000pc}
  E.~Braaten and J.~Lee,
  Nucl.\ Phys.\ B {\bf 586}, 427 (2000)
  [hep-ph/0004228].

\bibitem{Bodwin:2003wh}
  G.~T.~Bodwin and J.~Lee,
  Phys.\ Rev.\ D {\bf 69}, 054003 (2004)
  [hep-ph/0308016].

\bibitem{Sang:2009zz}
  W.~l.~Sang, L.~f.~Yang and Y.~q.~Chen,
  Phys.\ Rev.\ D {\bf 80}, 014013 (2009).

\bibitem{Bodwin:2012xc}
  G.~T.~Bodwin, U.~R.~Kim and J.~Lee,
  JHEP {\bf 1211}, 020 (2012)
  [arXiv:1208.5301 [hep-ph]].

\bibitem{Artoisenet:2014lpa}
  P.~Artoisenet and E.~Braaten,
  JHEP {\bf 1504}, 121 (2015)
  [arXiv:1412.3834 [hep-ph]].

\bibitem{Bodwin:2014bia}
  G.~T.~Bodwin, H.~S.~Chung, U.~R.~Kim and J.~Lee,
  Phys.\ Rev.\ D {\bf 91}, no. 7, 074013 (2015)
  [arXiv:1412.7106 [hep-ph]].

\bibitem{Gao:2016ihc}
  X.~Gao, Y.~Jia, L.~Li and X.~Xiong,
  Chin.\ Phys.\ C {\bf 41}, no. 2, 023103 (2017)
  [arXiv:1606.07455 [hep-ph]].

\bibitem{Sepahvand:2017gup}
  R.~Sepahvand and S.~Dadfar,
  Phys.\ Rev.\ D {\bf 95}, no. 3, 034012 (2017).

\bibitem{Zhang:2017xoj}
  P.~Zhang, Y.~Q.~Ma, Q.~Chen and K.~T.~Chao,
  Phys.\ Rev.\ D {\bf 96}, no. 9, 094016 (2017)
  [arXiv:1708.01129 [hep-ph]].

\bibitem{Feng:2017cjk}
  F.~Feng, S.~Ishaq, Y.~Jia and J.~Y.~Zhang,
  arXiv:1712.09986 [hep-ph].

\bibitem{Ma:2013yla}
  Y.~Q.~Ma, J.~W.~Qiu and H.~Zhang,
  Phys.\ Rev.\ D {\bf 89}, no. 9, 094029 (2014)
  [arXiv:1311.7078 [hep-ph]].

\bibitem{Hao:2009fa}
  G.~Hao, Y.~Zuo and C.~F.~Qiao,
  arXiv:0911.5539 [hep-ph].

\bibitem{Collins:1981uw}
  J.~C.~Collins and D.~E.~Soper,
  Nucl.\ Phys.\ B {\bf 194}, 445 (1982).

\bibitem{Nogueira:1991ex}
  P.~Nogueira,
  J.\ Comput.\ Phys.\  {\bf 105}, 279 (1993).

\bibitem{Petrelli:1997ge}
  A.~Petrelli, M.~Cacciari, M.~Greco, F.~Maltoni and M.~L.~Mangano,
  Nucl.\ Phys.\ B {\bf 514}, 245 (1998)
  [hep-ph/9707223].

\bibitem{Mertig:1990an}
  R.~Mertig, M.~Bohm and A.~Denner,
  Comput.\ Phys.\ Commun.\  {\bf 64}, 345 (1991).

\bibitem{Feng:2012tk}
  F.~Feng and R.~Mertig,
  arXiv:1212.3522 [hep-ph].

\bibitem{Feng:2012iq}
  F.~Feng,
  Comput.\ Phys.\ Commun.\  {\bf 183}, 2158 (2012)
  [arXiv:1204.2314 [hep-ph]].

\bibitem{Binoth:2000ps}
  T.~Binoth and G.~Heinrich,
  Nucl.\ Phys.\ B {\bf 585}, 741 (2000)
  [hep-ph/0004013].

\bibitem{Binoth:2003ak}
  T.~Binoth and G.~Heinrich,
  Nucl.\ Phys.\ B {\bf 680}, 375 (2004)
  doi:10.1016/j.nuclphysb.2003.12.023
  [hep-ph/0305234].

\bibitem{Smirnov:2013eza}
  A.~V.~Smirnov,
  Comput.\ Phys.\ Commun.\  {\bf 185}, 2090 (2014)
  doi:10.1016/j.cpc.2014.03.015
  [arXiv:1312.3186 [hep-ph]].

\bibitem{CubPack}
R.~Cools and A.~Haegemans,
ACM Trans. Math. Softw. {\bf 29} (2003), no.~3~287~C296.

\bibitem{parint}
http://www.cs.wmich.edu/parint, PARINT web site.

\bibitem{Feng:2015uha}
  F.~Feng, Y.~Jia and W.~L.~Sang,
  Phys.\ Rev.\ Lett.\  {\bf 115}, no. 22, 222001 (2015)
  [arXiv:1505.02665 [hep-ph]].

\bibitem{Feng:2017hlu}
  F.~Feng, Y.~Jia and W.~L.~Sang,
  Phys.\ Rev.\ Lett.\  {\bf 119}, 252001 (2017)
  [arXiv:1707.05758 [hep-ph]].

\bibitem{Artoisenet:2018dbs}
  P.~Artoisenet and E.~Braaten,
  arXiv:1810.02448 [hep-ph].

\bibitem{Zhang:2018}
  P.~Zhang, C.~Y.~Wang, X.~Liu, Y.~Q.~Ma, C.~Meng, K.~T.~Chao, to appear.


\end{thebibliography}
\end{document}